\documentclass[aps,prl,reprint,superscriptaddress,nofootinbib]{revtex4-1}
\pdfoutput=1
\usepackage{amsmath,amssymb,amsfonts,amsthm}
\usepackage{graphicx}
\usepackage{xcolor}
\usepackage{braket}
\usepackage{hyperref}

\newcommand{\bpm}{\begin{pmatrix}}
\newcommand{\epm}{\end{pmatrix}}

\def\0{{\sst{(0)}}}
\def\1{{\sst{(1)}}}
\def\2{{\sst{(2)}}}
\def\3{{\sst{(3)}}}
\def\4{{\sst{(4)}}}
\def\5{{\sst{(5)}}}
\def\6{{\sst{(6)}}}
\def\7{{\sst{(7)}}}
\def\8{{\sst{(8)}}}
\def\sst#1{{\scriptscriptstyle #1}}

\def\ie{\begin{equation}\begin{aligned}}
\def\fe{\end{aligned}\end{equation}}
\def\be{\begin{equation}}
\def\ee{\end{equation}}
\def\bea{\begin{eqnarray}}
\def\eea{\end{eqnarray}}
\def\ien{\begin{equation*}\begin{aligned}}
\def\fen{\end{aligned}\end{equation*}}

\newtheorem{conj}{Conjecture}

\begin{document}

\renewcommand\arraystretch{1.5}

\title{Signatures of Quantum Chaos in the D1D5 System}

\author{Haoyu Zhang}
\affiliation{Department of Physics and Astronomy, University of Southern California, Los Angeles, California 90089, USA}

\begin{abstract}
We investigate the emergence of random-matrix statistics in the D1D5 CFT by
studying second-order lifting matrices in low-energy near-BPS sectors. We
compare the $N=3$ finite-\(N\) lifting problems with the planar large-\(N\) limit at
fixed orbifold conformal weight and R charges. In the planar large-\(N\) limit at fixed orbifold energy, mixing between
single-cycle and multi-cycle states is suppressed, and the symmetry-resolved
lifting spectra display Poisson-like level statistics. At finite \(N\),
non-planar terms restore this mixing between different cycle structures. Within the resulting symmetry-resolved sectors, this finite-\(N\) mixing is
accompanied by level repulsion consistent with random-matrix behavior. These results suggest that, in the low-energy near-BPS sectors accessible to our
analysis, non-planar cycle-structure mixing at finite \(N\) is associated with
the onset of level repulsion and random-matrix-like spacing statistics.
\end{abstract}

\maketitle

\section{Introduction}
Recent developments in low-dimensional gravity have sharpened the meaning of
randomness in holography. In JT gravity, the gravitational path integral is
closely related to a matrix integral\cite{Saad:2019lba,Stanford:2019vob,Witten:2020wvy,Maxfield:2020ale,Mertens:2022irh,Turiaci:2023jfa}, suggesting an ensemble interpretation of
the boundary description. Related ideas have also appeared in three-dimensional
gravity, where simple gravitational path integrals may compute averaged CFT
data\cite{Cotler:2020ugk, Cotler:2020hgz, Belin:2023efa, Jafferis:2025yxt,Jafferis:2025jle, Jafferis:2025vyp}. These developments raise a complementary question in standard
string-theoretic holography. In examples such as the D1D5 system, the boundary
theory is expected to be a definite microscopic CFT, not an explicit ensemble of theories. How does the expected random-matrix behavior of chaotic
holographic systems arise from a fixed Hamiltonian\cite{Schlenker:2022dyo,Cotler:2022rud}?

The D1D5 CFT provides a useful setting in which to address this question, see also related discussions of random matrix statistics in \(\mathcal N=4\) SYM \cite{McLoughlin:2020zew, McLoughlin:2022jyt}. At the symmetric-product orbifold point, \(\mathrm{Sym}^N(T^4)\), the theory
is exactly solvable\cite{Dijkgraaf:1996xw,David:2002wn} and contains large degenerate subspaces of states. Moving away from the
orbifold point generically lifts these degeneracies\cite{Gava:2002xb, Avery:2010er, Avery:2010hs, Guo:2019ady,Guo:2020gxm, Gaberdiel:2023lco}. At second order in
conformal perturbation theory, the lifting is governed by finite-dimensional
matrices acting within the degenerate orbifold subspaces. These finite-dimensional second-order lifting matrices, which we refer to
throughout the paper as the deformed Hamiltonian, can be written as
cohomological Laplacians built from the first-order action of the deformation
supercharge. We study their level statistics as a diagnostic
of integrable versus chaotic behavior\cite{Berry:1977,Bohigas:1984,Guhr:1998,Haake:2010}.

The lifting matrices of orbifold $1/4$ BPS states in the D1D5 CFT have previously been studied at \(N=2\)
and in the planar large-\(N\) limit\cite{Guo:2019ady,Guo:2020gxm, Gaberdiel:2023lco}. Here we compare two complementary regimes:
the planar large-\(N\) regime and the genuinely finite-\(N\) regime. Planar integrability at large single-cycle length was discussed in \cite{Gaberdiel:2023lco,Frolov:2023pjw,Gaberdiel:2024nge}. We analyze
nearest-neighbor level-spacing distributions in both cases. We find evidence that in the fixed-energy planar large-\(N\) limit, even at
finite twist \(w\), the lifting spectrum retains an approximately integrable,
Poisson-like organization. By contrast, at finite \(N\), sectors
that are distinguished in the planar limit can mix. In particular, single-cycle
and multi-cycle states can enter the same lifting problem. This finite-\(N\) mixing is accompanied by level repulsion and spacing
statistics closer to the random-matrix benchmark.\footnote{While this work was nearing completion, we became aware of related work
discussing non-planar corrections to the planar D1D5 lifting problem and the
emergence of random-matrix statistics \cite{Gaberdiel:2026jor}.}
\section{The deformed Hamiltonian and the supercharge complex}
\label{sec:deformed-hamiltonian}

The common object in both the finite-\(N\) and planar large-\(N\) analyses is
the second-order deformed Hamiltonian acting on a degenerate orbifold subspace.
At the orbifold point we focus on states that are right-moving BPS, while
allowing nontrivial left-moving excitations. Moving away from the orbifold
point by an exactly marginal deformation in the twisted sector lifts these
degeneracies\cite{Gava:2002xb, Avery:2010er, Avery:2010hs, Guo:2019ady,Guo:2020gxm, Gaberdiel:2023lco}.
To second order in conformal perturbation theory, the deformed Hamiltonian is
determined by the first-order action of the deformation supercharge,\footnote{Equivalently, with the standard convention in which the eigenvalues give the
shift of the total conformal dimension,
\[
\delta(h+\tilde h)=2\{Q,Q^\dagger\},
\qquad
\delta h=\delta\tilde h .
\]
The equality \(\delta h=\delta\tilde h\) follows from locality of the
deformation, which preserves the Lorentz spin \(h-\tilde h\).}
\[
\Delta=\{Q,Q^\dagger\}.
\]

Following \cite{Chang:2025rqy}, we use the component of the deformation
supercharge that raises the right-moving \(R\)-spin by one half, while preserving the left-moving quantum numbers
such as \(h\) and \(j\). It is therefore natural to introduce the cochain degree
\[
p=2\tilde j .
\]
For each fixed pair \((h,j)\), the deformed supercharge defines a finite-dimensional cochain complex
\[
\cdots \longrightarrow C^p_{h,j}
\xrightarrow{Q_p}
C^{p+1}_{h,j}
\longrightarrow \cdots ,
\]
by the nilpotency condition $Q_{p+1}Q_p=0$. In the following, we
will usually suppress the fixed labels \((h,j)\) and write simply \(C^p\).

The deformed Hamiltonian\(\Delta\)
preserves the cochain degree \(p\). Thus each space \(C^p\) defines a block of the deformed
Hamiltonian. On this block, the second-order lifting operator is the
cohomological Laplacian
\[
\Delta_p
=
Q_{p-1}Q_{p-1}^{\dagger}
+
Q_p^{\dagger}Q_p ,
\]
with the understanding that maps outside the complex are set to zero. The two terms in \(\Delta_p\) have orthogonal support since the image of \(Q_{p-1}\) lies in the kernel of \(Q_p\), 
\[ Q_{p-1}Q_{p-1}^\dagger,\qquad Q_p^\dagger Q_p
\]
act on orthogonal components of \(C^p\). In other words, the \(Q_{p-1}\)-exact subspace \(\operatorname{im}Q_{p-1}\) is orthogonal to
the \(Q_p^\dagger\)-exact subspace \(\operatorname{im}Q_p^\dagger\). Consequently, the nonzero spectrum of
\(\Delta_p\) is the multiset union of the singular-value contributions from
the adjacent maps \(Q_{p-1}\) and \(Q_p\).

\section{Deformed Hamiltonian at finite \(N\)}
In the
\(N=3\) sector considered below, the supercharge cochain complex is
\[
\begin{aligned}
0&\to V_{(1^3),0}^{0}
\xrightarrow{Q_0} V_{(1,2),0}^{1}
\xrightarrow{Q_1} V_{(1^3),0}^{2}\oplus V_{(3),0}^{2} \\
&\xrightarrow{Q_2} V_{(1,2),0}^{3}
\xrightarrow{Q_3} V_{(1^3),0}^{4}
\to 0 .
\end{aligned}
\]
Here the superscript denotes the cochain degree \(p\), which is related to the
right-moving \(R\) charge by \(p=2\tilde j\).\footnote{The complex admits a reflection symmetry
\[
\star_p:C^p\longrightarrow C^{4-p},
\qquad
\star_p\Delta_p\star_p^{-1}
=
\Delta_{4-p}.
\]
The relation \(C^1\leftrightarrow C^3\) follows from long-multiplet
recombination. The endpoint relation \(C^0\leftrightarrow C^4\) can be
understood by reducing the relevant \(N=3\) matrix elements to the lifting
problem at \(N=2\), where there is only one nontrivial long-multiplet
recombination structure.
} The first subscript specifies the
cycle shape,
\[
(1^{n_1},2^{n_2},3^{n_3},\ldots),
\qquad
\sum_k k n_k=N,
\]
or equivalently a partition of \(N\). This partition labels a conjugacy class
of \(S_N\), and hence the corresponding twisted sector of the symmetric-product
orbifold Hilbert space. The second subscript labels the representation of the outer automorphism group
\(SU(2)_{\mathrm{outer}}\); see appendix A of \cite{Chang:2025wgo} for further details.

We will focus on the space
\[
C^1
=
V_{(1,2),0}^{1},
\qquad
\Delta_1
=
Q_0Q_0^\dagger+Q_1^\dagger Q_1 .
\]
The resulting lifted states should be interpreted in the
context of long-multiplet recombination. Short
multiplets can combine into a long multiplet with right-moving \(R\)-charge
content \(1_{\tilde j}\oplus2_{\tilde j+\frac12}\oplus1_{\tilde j+1}\).
The supersymmetry algebra then implies that all four components share the same
anomalous dimension. Therefore the spectrum obtained from \(V_{(1,2),0}^{1}\)
represents one block of the full long multiplet structure. Part of the spectrum
of \(\Delta_1\) is inherited from the adjacent lifting problem in
\(V_{(1^3),0}^{0}\).
When extracting the genuinely new eigenvalues in this \(C^1\) block, we remove
these inherited eigenvalues. This prevents double-counting
the same recombination channel in adjacent cochain degrees.

\section{Deformed Hamiltonian in the planar large-\(N\) limit}
We now compare the finite-\(N\) lifting problem with its planar large-\(N\)
limit. The relevant limit keeps the orbifold energy fixed, \(h=O(1)\), while
taking \(N\to\infty\). In this regime, only finitely many copies are excited,
while the number of spectator copies with no excitations remains of order \(N\). The leading contribution to the deformed supercharge is the part in which an active cycle joins with, or splits off, an
unexcited length-one cycle. Processes involving two active cycles, or splitting
one active cycle into two active cycles, are suppressed by inverse powers of
\(N\). We therefore write
\[
Q = Q_\infty + Q_{\rm np},
\]
where \(Q_\infty\) denotes the planar part of the supercharge and
\(Q_{\rm np}\) contains the non-planar corrections. Keeping only \(Q_\infty\),
the supercharge complex reduces to a planar complex of single-cycle states,
\[
\begin{aligned}
0&\to V_{(1)}^{0}
\xrightarrow{Q_{\infty,0}} V_{(2)}^{1}
\xrightarrow{Q_{\infty,1}} V_{(3)}^{2}\oplus V_{(1)}^{2}\\
&\xrightarrow{Q_{\infty,2}} V_{(4)}^{3}\oplus V_{(2)}^{3}
\xrightarrow{Q_{\infty,3}}\cdots .
\end{aligned}
\]
Here the superscript denotes the cochain degree, while the subscript records the cycle length of the active cycle; see the
\hyperref[app:largeN-spectra]{appendix discussion of the planar large-\(N\) single-cycle complex}
for more details.
The corresponding deformed Hamiltonian for each space is
\[
\Delta^{(\infty)}_p
=
Q_{\infty,p-1}Q_{\infty,p-1}^\dagger
+
Q_{\infty,p}^\dagger Q_{\infty,p}.
\]
We focus on the lifting problem in the space \(V^1_{(2)}\). Part of the
spectrum of \(\Delta^{(\infty)}_1\) is inherited from the adjacent lifting
problem in \(V^0_{(1)}\), and we therefore remove these inherited values.

\section{Nearest-neighbor level-spacing distributions}
\subsubsection*{Symmetry resolution and primary projection}
Before applying spectral-statistics diagnostics, we resolve the exact
symmetries of the lifting problem. The lifting operator commutes with the
conserved charges that label the orbifold sector and, in particular, with the
Casimirs of \(SU(2)_a\) and \(SU(2)_b\).\footnote{These two
\(SU(2)\) symmetries are not R-symmetries. They are preserved by the exactly
marginal deformation as well; see \cite{Gaberdiel:2024nge,Chang:2025wgo} for further
details.} We therefore decompose each cochain
space into symmetry-resolved blocks,
\[
C^p
=
\bigoplus_{h,j,c_1,c_2}
C^p_{h,j,c_1,c_2},
\]
where \(h\) is the orbifold conformal weight, \(j\) is the relevant
\(SU(2)\) R charge, and \(c_1,c_2\) are the \(SU(2)_a\) and \(SU(2)_b\) Casimir
eigenvalues. We further study nearest-neighbor level-spacing distributions after
removing descendants. The precise notion of primarity
depends on the regime under consideration. At finite \(N\), we restrict to superconformal primaries of the finite-\(N\) contracted large
\(\mathcal N=(4,4)\) algebra. In
the planar large-\(N\) limit, we instead restrict to primaries of the global
\(SU(1,1|2)\) subalgebra of the contracted large \(\mathcal N=(4,4)\) algebra.

Although the orbifold theory contains large degenerate subspaces, the number
of independent levels becomes much smaller after resolving the exact symmetry
sectors and imposing the appropriate primary conditions. In particular, after
fixing the \(SU(2)_a\times SU(2)_b\) Casimir sector, many low-lying sectors contain too few levels for a reliable statistical analysis; examples are provided in the
\hyperref[app:finiteN-spectra]{finite-\(N\)} and
\hyperref[app:largeN-spectra]{planar large-\(N\)} appendix tables. We therefore
present in the main text only those sectors for which the available data are
sufficient to illustrate the qualitative behavior of the level-spacing
distribution.
\subsubsection*{Unfolding }

Following Ref.~\cite{McLoughlin:2020zew}, we unfold each
symmetry-resolved Casimir block independently before computing nearest-neighbor
spacings. The unfolded spacings are then pooled to form the histograms. We compare the unfolded spacing distribution with the Poisson
distribution $P_{\rm Poisson}(s)=e^{-s}$ and the GOE Wigner surmise $
P_{\rm GOE}(s)=\frac{\pi}{2}s\,e^{-\pi s^2/4}.
$ These are the standard benchmarks for integrable and time-reversal-invariant
chaotic spectra \cite{Berry:1977,Bohigas:1984,Mehta:2004,Haake:2010}. The precise Casimir sectors used
in each histogram are listed in the
\hyperref[app:spacing-data]{appendix section on Casimir sectors}. 
\subsubsection*{Results}
\begin{figure*}[t]
    \centering
    \includegraphics[width=0.92\textwidth]{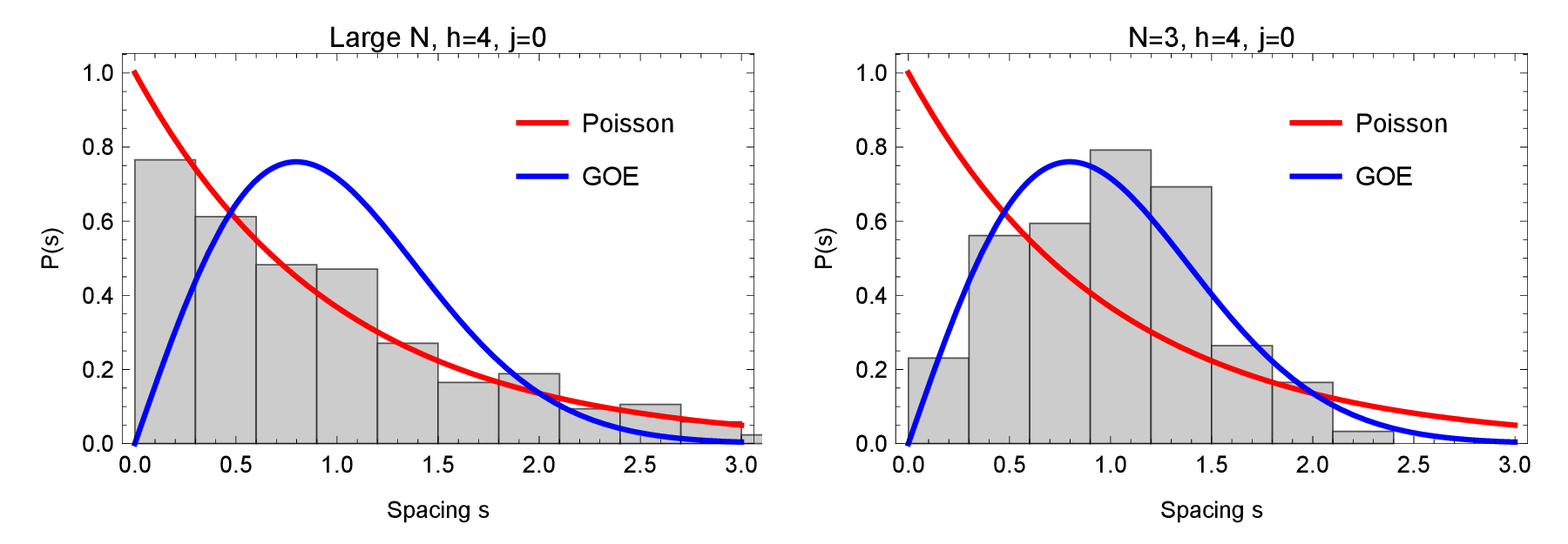}
    \vspace{0.6em}
    \includegraphics[width=0.92\textwidth]{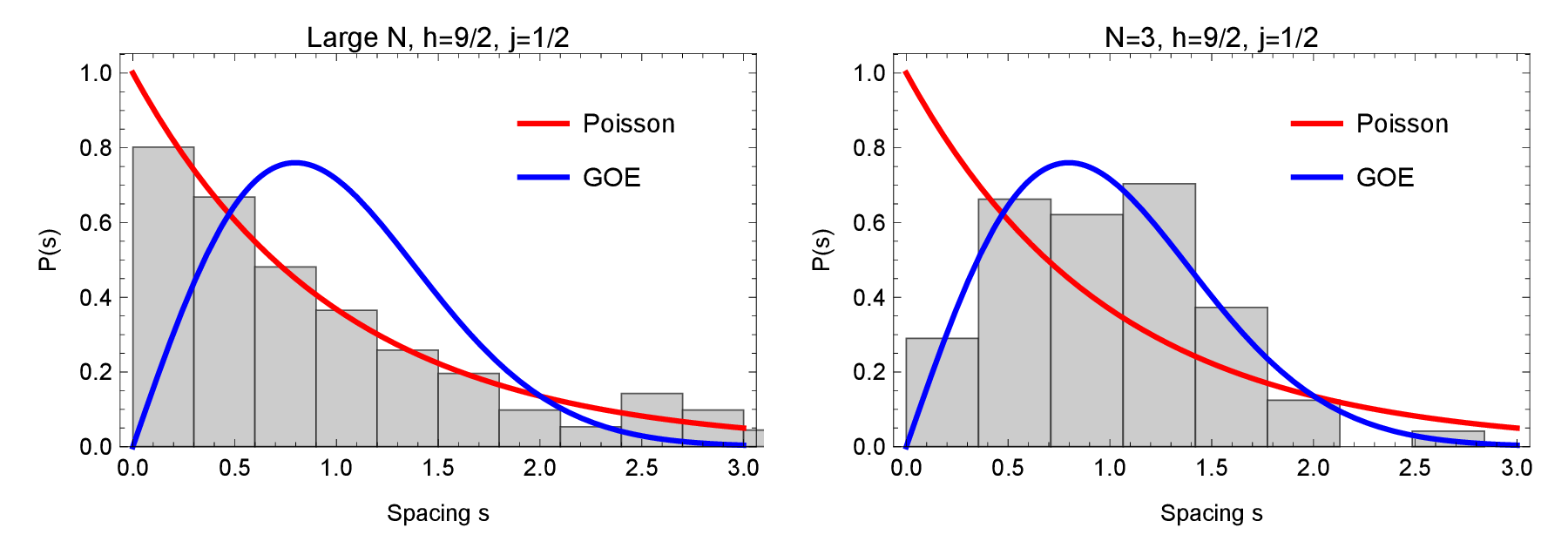}
    \caption{Nearest-neighbor spacing distributions at
    \((h,j,\tilde j)=(4,0,1/2)\) (top) and
    \((h,j,\tilde j)=(9/2,1/2,1/2)\) (bottom). Each row compares
    the planar large-\(N\) spectrum with the finite-\(N\), \(N=3\), spectrum
    after symmetry resolution and unfolding.}
    \label{fig:spacing-transition}
\end{figure*}

We have analyzed spectra with \(0\leq h\leq \frac{9}{2}\). In both the finite-\(N\)
and planar large-\(N\) computations, the lifting matrices are determined up to
an overall universal normalization factor.\footnote{
Since the spectra are unfolded before nearest-neighbor spacings are computed,
this overall normalization does not affect the spacing distributions. The
normalization can be fixed using the standard covering-space normalization of
twist correlators, following
\cite{Lunin:2000yv,Lunin:2001pw,Dei:2019iym}.
} The first sectors in our data set
with enough levels to display qualitative spectral-statistical behavior occur
at \(h=4\) and \(h=\frac{9}{2}\). For the sectors displayed below, the relevant
\(R\)-charge is \(j=0\) when \(h\) is integer and \(j=\frac12\) when \(h\)
is half-integer. The resulting comparison
between the planar large-\(N\) and finite-\(N\) spacing distributions is shown
in Fig.~\ref{fig:spacing-transition}.
In both examples, the planar large-\(N\) spectra retain substantial weight near
small spacings and are closer to Poisson statistics, while the finite-\(N\)
spectra show a visible suppression of small spacings and a shift toward the
GOE curve, indicating level repulsion. In all sectors studied, after removing descendants and resolving the residual \( SU(2)_a\times SU(2)_b\)  symmetry, we find no accidental degeneracies among positive lifting eigenvalues of primary states. This motivates the following conjecture:
\begin{conj}
After resolving the spectrum into irreducible representations of the residual
\(  SU(2)_a\times SU(2)_b\) symmetry, the exactly marginal
deformation generically removes all degeneracies among positive lifting eigenvalues of
orbifold primary states.
\end{conj}
\section{Discussion}
Our results reveal a crossover in the spectral statistics of near-BPS states
between the planar large-\(N\) regime and the genuinely finite-\(N\) regime.
At large \(N\), after resolving the spectrum by the relevant conserved quantum
numbers, the deformed Hamiltonian retains an approximately integrable organization,
with level spacings closer to Poisson statistics. At finite \(N\), however,
states that would be distinguished in the planar limit can mix. In particular,
finite-\(N\) effects allow single-cycle and multi-cycle states to participate in
the same lifting problem. The emergence of random-matrix statistics in the D1D5
CFT is therefore not merely a consequence of having a large Hilbert space. The data suggest that finite-N cycle-structure mixing is an important mechanism behind the observed level repulsion. It would be interesting to extend this analysis
to genuinely non-BPS states at the orbifold point, where the relation between random-matrix statistics
and integrability has recently been explored in \cite{Fabri:2025rok}.

A natural next step is to follow sectors with fixed orbifold energy while
varying \(N\), as in \cite{Budzik:2023vtr,Chang:2024lxt}. This would directly
test how the finite-\(N\) spectrum approaches the fixed-energy planar regime.
One expects an interpolation at fixed \(h=O(1)\): at small \(N\), finite-\(N\)
mixing can produce level repulsion and random-matrix-like statistics, while as
\(N\) is increased the same sector should become increasingly planar and hence
more Poisson-like. This interpolation is different from the approach to the
black-hole regime, where the relevant charges scale with \(N\) \cite{Strominger:1996sh,David:2002wn, Kraus:2006wn} and the
suppression of non-planar mixing may break down. Establishing
this crossover would provide a direct quantitative test of the role of
finite-\(N\) dynamics in the onset of spectral chaos.

It would also be interesting to connect the onset of random-matrix statistics
with the notion of fortuity
\cite{Chang:2024lxt,Chang:2024zqi,Chang:2025rqy,Choi:2025bhi,Kim:2025vup, Belin:2025hsg,Hughes:2025car,
Chen:2025sum}
in the BPS spectrum. Fortuitous BPS states are BPS due to finite-\(N\) effects and do not persist
in the planar limit, so their existence is intrinsically tied to the
finite-\(N\) structure of the Hilbert space. Random-matrix behavior in the BPS spectrum has been proposed
\cite{Lin:2022rzw,Lin:2022zxd} in the language of \(\mathcal{N}=2\) JT gravity.
Although the present work concerns near-BPS states that are lifted after deforming
away from the orbifold point, rather than exact BPS states themselves, it would be interesting to test whether these two phenomena are related through the chaos invasion mechanism proposed in
\cite{Chen:2024oqv}. Related diagnostics based on the Berry curvature of BPS
chaos were studied in \cite{Chen:2026vml}.

\section*{Acknowledgments}
We thank Yiming Chen for collaboration at an early stage of this work and for
many valuable insights at various stages of this project. We also thank Chi-Ming Chang and Ying-Hsuan Lin for
collaboration on related D1D5 projects. We benefited from discussions with
Nathan Benjamin, Chi-Ming Chang, Matthias Gaberdiel, Bin Guo, Maciej Kolanowski, Ho-Tat Lam, 
Ji-Hoon Lee, Wei Li, Yue-Zhou Li, Henry Lin, Ying-Hsuan Lin, Juan Maldacena, Joaquin Turiaci, Dian-Dian Wang, and Nick
Warner. Parts of this work were presented at UCLA, Cooks Branch, USC, and Princeton,
and I thank the participants for useful comments and discussions. The author
also thanks Codex for PRL formatting and arXiv source-packaging assistance. HZ is supported by the U.S. Department of Energy, Office of Science, under grant Contract Number DE-SC0026324.

\onecolumngrid
\appendix
\section{Characters and descendant counting}
The descendant content of a primary with fixed quantum numbers provides an
important consistency check on the spectrum. In the character decomposition,
the coefficient of a given term $q^hy^{2j}$ counts the degeneracy of descendants with quantum
numbers \((h,2j)\), where \(h\) is the holomorphic conformal weight and \(j\)
is the left-moving \(SU(2)_R\) spin.

At \(N=3\), the relevant contracted large $\mathcal{N}=(4,4)$
characters \cite{Eguchi:1987sm, Eguchi:1987wf} are
\bea
&&\chi_{h,j=0} = q^h +q^{h+\frac{1}{2}} (4 y^{-1}+4) + q^{h+1} (22+ 7 y^{-2}+ 7y^2) + q^{h+\frac{3}{2}} (8 y^{-3}+56 y^{-1}+56 y +8y^3)+\cdots\crcr
&&\chi_{h,j=\frac{1}{2}} = q^{h}(y+y^{-1}) +q^{h+\frac{1}{2}} (8+ 4 y^{-2}+4 y^2) + q^{h+1} (6 y^{-3}+ 28 y^{-1}+28 y + 6y^3) +\cdots \label{eq:finiteN-characters}
\eea
In the planar large-\(N\) limit, the relevant global $SU(1,1|2)$ long characters, see for example in \cite{Hughes:2026naj}, are
\bea
&&\chi_{h,j=0} = q^h + q^{h+\frac{1}{2}}(2y +2y^{-1}) + q^{h+1}(5+y^{-2}+y^2)+ q^{h+\frac{3}{2}}(4y^{-1}+4y)+\cdots\crcr
&&\chi_{h,j=\frac{1}{2}} = q^{h}(y+y^{-1})+q^{h+\frac{1}{2}}(4+ 2y^{-2}+2y^2)+q^{h+1}(y^{-3}+6y^{-1}+6y+y^3)+\cdots \label{eq:largeN-characters}
\eea
These character expansions are used below to identify which lifting
eigenvalues are inherited from lower-lying primaries as descendant
contributions.

\section{Lifting spectra and descendant checks}

In this appendix we collect lifting spectra used as
consistency checks for the primary projection and descendant subtraction. We
list only the nonzero lifting eigenvalues, omitting states that remain BPS
after the deformation. Throughout this appendix, \(n\times \lambda\) denotes
an eigenvalue \(\lambda\) with multiplicity \(n\); when the multiplicity is
one, we simply write \(\lambda\). When comparing with character multiplicities, the multiplicity in a fixed
Casimir sector must be weighted by the dimension of the corresponding
\( SU(2)_b\times SU(2)_a\) representation. Thus, if
\(c_i=s_i(s_i+1)\), an entry \(n\times\lambda\) in the sector
\((c_1,c_2)\) contributes
\[
n(2s_1+1)(2s_2+1)
\]
states with lifting eigenvalue \(\lambda\).
\subsection{Finite-\(N\) spectra in the \(N=3\) complex}\label{app:finiteN-spectra}
\subsubsection{Spectrum in \(V^0_{(1^3),0}\)}

At \(h=1\), \(j=0\), there is one distinct nonzero anomalous dimension. The
decomposition by \( SU(2)_b\times SU(2)_a\) Casimir sector is
\[
\begin{array}{c|c}
\text{Casimir sector} & \text{Lifting eigenvalues} \\
\hline
(2,0) & \left\{\frac{1}{2}\right\}
\end{array}
\]

At \(h=\frac{3}{2}\), \(j=\frac{1}{2}\), there are two distinct nonzero
anomalous dimensions. The sector-by-sector decomposition is
\[
\begin{array}{c|c}
\text{Casimir sector} & \text{Lifting eigenvalues} \\
\hline
(\frac{3}{4},0) & \left\{\frac{1}{2},\frac{3}{8}\right\} \\
(2,\frac{3}{4}) & \left\{\frac{1}{2}\right\} \\
(\frac{15}{4},0) & \left\{\frac{1}{2}\right\}
\end{array}
\]
The eigenvalue \(\frac{3}{8}\) accounts for two states, corresponding to the
two primaries at the orbifold point. The eigenvalue \(\frac{1}{2}\) has total
multiplicity \(12\), which is accounted for by the four descendants of each
\(h=1\), \(j=0\) primary, in agreement with the character expansion
\eqref{eq:finiteN-characters}.

At \(h=2\), \(j=0\), there are six distinct nonzero anomalous dimensions. The
decomposition by Casimir sector is
\[
\begin{array}{c|c}
\text{Casimir sector} & \text{Lifting eigenvalues} \\
\hline
(0,0)
& \left\{0.951444\ldots,\;0.517306\ldots,\;\frac{1}{2},\;
2\times\frac{3}{8}\right\} \\

(\frac{3}{4},\frac{3}{4})
& \left\{3\times\frac{1}{2},\;2\times\frac{3}{8}\right\} \\

(\frac{15}{4},\frac{3}{4})
& \left\{\frac{3}{4},\;3\times\frac{1}{2}\right\} \\

(2,0)
& \left\{5\times\frac{1}{2},\;2\times\frac{3}{8}\right\} \\

(2,2)
& \left\{\frac{1}{2}\right\} \\

(6,0)
& \left\{1,\;\frac{1}{2}\right\}
\end{array}
\]
After subtracting the descendant contributions associated with the lower-level
eigenvalues \(\frac{1}{2}\) and \(\frac{3}{8}\), the remaining positive
eigenvalues account for \(15\) states. This agrees with the number of primaries
at \(h=2\), \(j=0\). More explicitly, the total multiplicities
\[
66=3\times 22,
\qquad
16=2\times 8
\]
of the inherited eigenvalues \(\frac{1}{2}\) and \(\frac{3}{8}\), respectively,
are reproduced by the descendant multiplicities in
\eqref{eq:finiteN-characters}, after including the dimensions of the
\(SU(2)_a\times SU(2)_b\) representations.

\subsubsection{Spectrum in \(V^1_{(1,2),0}\)}

At \(h=1\), \(j=0\), the nonzero lifting spectrum decomposes as
\[
\begin{array}{c|c}
\text{Casimir sector} & \text{Lifting eigenvalues} \\
\hline
(\frac{3}{4},0) & \left\{\frac{4}{9}\right\} \\
(2,\frac{3}{4}) & \left\{\frac{1}{2}\right\}
\end{array}
\]
The eigenvalue \(\frac{1}{2}\) is inherited from the lifting spectrum in
\(V^0_{(1^3),0}\).

At \(h=\frac{3}{2}\), \(j=\frac{1}{2}\), the nonzero lifting spectrum is
\[
\begin{array}{c|c}
\text{Casimir sector} & \text{Lifting eigenvalues} \\
\hline
(0,0) & \left\{\frac{4}{9},\;\frac{10}{27}\right\} \\
(2,0) & \left\{\frac{2}{3},\;\frac{1}{2},\;\frac{4}{9}\right\} \\
(2,2) & \left\{\frac{1}{2}\right\} \\
(\frac{3}{4},\frac{3}{4})
& \left\{\frac{1}{2},\;\frac{4}{9},\;\frac{3}{8}\right\} \\
(\frac{15}{4},\frac{3}{4})
& \left\{\frac{1}{2},\;\frac{2}{3}\right\}
\end{array}
\]
The multiplicities of the inherited modes agree with the descendant structure
of the lower-lying primaries, as determined from
\eqref{eq:finiteN-characters}. In particular, the eigenvalues 
\be
\frac{1}{2},\frac{3}{8}
\ee
can be traced back to the lifting spectrum in \(V^0_{(1^3),0}\).

At \(h=2\), \(j=0\), the nonzero lifting spectrum is
\[
\begin{array}{c|c}
\text{Casimir sector} & \text{Lifting eigenvalues} \\
\hline
(0,\frac{3}{4})
& \left\{
2\times\frac{10}{27},\;
2\times\frac{3}{8},\;
3\times\frac{4}{9},\;
\frac{1}{2},\;
0.5173\ldots,\;
0.95144\ldots
\right\} \\

(\frac{3}{4},0)
& \left\{
0.307559\ldots,\;
2\times\frac{10}{27},\;
2\times\frac{3}{8},\;
5\times\frac{4}{9},\;
3\times\frac{1}{2},\;
0.539881\ldots,\;
2\times\frac{2}{3},\;
0.979111\ldots
\right\} \\

(\frac{3}{4},2)
& \left\{
2\times\frac{3}{8},\;
\frac{4}{9},\;
3\times\frac{1}{2},\;
\frac{8}{9}
\right\} \\

(2,\frac{3}{4})
& \left\{
0.214866\ldots,\;
0.325125\ldots,\;
2\times\frac{3}{8},\;
3\times\frac{4}{9},\;
6\times\frac{1}{2},\;
4\times\frac{2}{3},\;
1.117073\ldots
\right\} \\

(\frac{15}{4},0)
& \left\{
\frac{4}{9},\;
3\times\frac{1}{2},\;
4\times\frac{2}{3},\;
\frac{3}{4}
\right\} \\

(2,\frac{15}{4})
& \left\{\frac{1}{2}\right\} \\

(\frac{15}{4},2)
& \left\{
3\times\frac{1}{2},\;
2\times\frac{2}{3},\;
\frac{3}{4}
\right\} \\

(6,\frac{3}{4})
& \left\{
\frac{1}{2},\;
2\times\frac{2}{3},\;
1
\right\} \\

(6,\frac{15}{4})
& \left\{0.296296\ldots\right\}
\end{array}
\]
The multiplicities of the inherited modes again match the descendant structure
of the lower-lying primaries. In particular, the eigenvalues
\[
1,\qquad
0.5173\ldots,\qquad
\frac{3}{4},\qquad
0.95144\ldots,\qquad
\frac{1}{2},\qquad
\frac{3}{8}
\]
can be traced to the lifting spectrum in \(V^0_{(1^3),0}\). This provides a nontrivial
consistency check on the descendant subtraction used in the primary projection.

\subsection{Planar large-\(N\) single-cycle complex}\label{app:largeN-spectra}
As discussed in \cite{Lunin:2001pw}, an orbifold right-moving BPS state with a single active cycle of
length \(w\) can carry four possible right-moving \(R\)-charges,
\[
\tilde j \in
\left\{
\frac{w-1}{2},\,
\frac{w}{2},\,
\frac{w}{2},\,
\frac{w+1}{2}
\right\}.
\]
Equivalently, the middle value \(\tilde j=w/2\) occurs with multiplicity two. In the main text, we denote by
\[
V_{(w)}^{p}, \qquad p=2\tilde j ,
\]
the space of states with one active cycle of length \(w\) and right-moving
\(R\)-charge \(\tilde j=p/2\). Thus the superscript \(p\) labels the cochain
degree, while the subscript \((w)\) records the single-cycle structure.

\subsubsection{Planar spectra in \(V^0_{(1)}\)}
For \(h=1\) and \(j=0\), the lifting spectrum decomposes by Casimir sector as
\[
\begin{array}{c|c}
\text{Casimir sector} & \text{Lifting eigenvalues} \\
\hline
(2,0) & \left\{\tfrac{2}{3}\right\} 
\end{array}
\]
For \(h=\frac{3}{2}\) and \(j=\frac{1}{2}\), the lifting spectrum decomposes by Casimir sector as
\[
\begin{array}{c|c}
\text{Casimir sector} & \text{Lifting eigenvalues} \\
\hline
(2,\tfrac{3}{4}) & \left\{\tfrac{2}{3}\right\}\\
(\tfrac{3}{4},0) & \left\{\tfrac{1}{2}\right\}  
\end{array}
\]
At \(h=\frac{3}{2}\), a primary at \(h=1\), \(j=0\) has two global
descendants with \(j=\frac{1}{2}\). Therefore the eigenvalue
\(\tfrac{2}{3}\), inherited from the \(h=1\) primary spectrum, is expected to
appear with total multiplicity
\[
2\times 3 = 6 .
\]
This agrees with the sector decomposition above, since the entry
\(\tfrac{2}{3}\) in the sector \((2,\tfrac{3}{4})\) contributes
\[
1\times 3\times 2 = 6
\]
states. 
For \(h=2\) and \(j=0\), the lifting spectrum decomposes by Casimir sector as
\[
\begin{array}{c|c}
\text{Casimir sector} & \text{Lifting eigenvalues} \\
\hline
(0,0) & \left\{1,\frac{3}{8}\right\}\\
(\tfrac{3}{4},\tfrac{3}{4}) & \left\{2\times \tfrac{1}{2}\right\}  \\
(2,2) & \left\{ \tfrac{2}{3}\right\}\\
(\frac{15}{4},\frac{3}{4}) & \left\{ 1\right\}\\
(2,0) & \left\{ 2\times\frac{2}{3}\right\}
\end{array}
\]
At \(h=2\), the descendants of the \(h=1\), \(j=0\) primaries give another
consistency check. For example, a primary at \(h=1\), \(j=0\) has five global descendants
at \(h=2\), \(j=0\). Since there are three primary states with lifting
eigenvalue \(\tfrac{2}{3}\) at \(h=1\), we expect
\[
5\times 3 = 15
\]
descendants with the same lifting eigenvalue at \(h=2\). This agrees with the
sector-by-sector decomposition, where the eigenvalue \(\tfrac{2}{3}\) appears
in the sectors \((2,2)\) and \((2,0)\), giving total multiplicity
\[
1\times 3\times 3 + 2\times 3\times 1 = 15 .
\]
Similarly, the new primary states at \(h=\frac{3}{2}\), \(j=\frac{1}{2}\),
with lifting eigenvalue \(\tfrac{1}{2}\), have four global descendants at
\(h=2\), \(j=0\). Since the eigenvalue \(\tfrac{1}{2}\) accounts for
\[
1\times 2\times 1=2
\]
primary states at \(h=\frac{3}{2}\), we expect
\[
4\times 2=8
\]
descendants with the same lifting eigenvalue at \(h=2\). This agrees with
the entry \(2\times \tfrac{1}{2}\) in the sector
\((\tfrac{3}{4},\tfrac{3}{4})\).

\subsubsection{Planar spectra in \(V^1_{(2)}\)}
For \(h=1\) and \(j=0\), the lifting spectrum decomposes by Casimir sector as
\[
\begin{array}{c|c}
\text{Casimir sector} & \text{Lifting eigenvalues} \\
\hline
(2,\tfrac{3}{4}) & \left\{\tfrac{2}{3}\right\} \\
(\tfrac{3}{4},0) & \left\{\tfrac{16}{81}\right\}
\end{array}
\]

For \(h=\tfrac{3}{2}\) and \(j=\tfrac{1}{2}\), the corresponding decomposition is
\[
\begin{array}{c|c}
\text{Casimir sector} & \text{Lifting eigenvalues} \\
\hline
(0,0) & \left\{\tfrac{8}{27},\,\tfrac{32}{243}\,\right\} \\
(2,0) & \left\{\tfrac{2}{3},\,\tfrac{40}{243}\right\} \\
(2,2) & \left\{\tfrac{2}{3}\right\} \\
(\tfrac{15}{4},\tfrac{3}{4}) & \left\{\tfrac{8}{27}\right\} \\
(\tfrac{3}{4},\tfrac{3}{4}) & \left\{\tfrac{1}{2},\,\tfrac{16}{81}\right\}
\end{array}
\]
The eigenvalues \(\frac{2}{3}\) and \(\frac{1}{2}\)
are inherited from the deformed Hamiltonian on \(V_{(1)}^0\). This spectrum is consistent with the $SU(1,1|2)$ descendant structure from the
\(h=1\), \(j=0\) primaries, cf. Eq.~\eqref{eq:largeN-characters}. A primary at \(h=1\), \(j=0\) has two global
descendants at \(h=\tfrac{3}{2}\), \(j=\tfrac{1}{2}\). In particular, the
eigenvalue \(\tfrac{2}{3}\) appearing at \(h=\tfrac{3}{2}\),
\(j=\tfrac{1}{2}\) can be identified with the global descendants of the
\(h=1\), \(j=0\) primaries with the same lifting eigenvalue. Since there are
six such primaries at \(h=1\), we expect
\[
2 \times 6 = 12
\]
corresponding descendants at \(h=\tfrac{3}{2}\), \(j=\tfrac{1}{2}\). This
agrees with the multiplicity
\[
3 + 3 \times 3 = 12
\]
found in the sector decomposition above. The same reasoning applies to the inherited
eigenvalue \(\tfrac{16}{81}\).
For \(h=2\) and \(j=0\), one finds
\[
\begin{array}{c|c}
\text{Casimir sector} & \text{Lifting eigenvalues} \\
\hline
(0,\tfrac{3}{4}) 
& \left\{1,\,\tfrac{3}{8},\,2\times\tfrac{8}{27},\,2\times\tfrac{32}{243} \right\} \\

(\tfrac{3}{4},0) 
& \left\{2\times\tfrac{1}{2},\,\tfrac{32}{81},\,0.3078\ldots,\,0.2598\ldots,\,0.2148\ldots,\,
2\times\tfrac{16}{81},\,0.1279\ldots,\,0.0865\ldots,\,\right\} \\

(\tfrac{3}{4},2) 
& \left\{2\times\tfrac{1}{2},\,\tfrac{32}{81},\,\tfrac{16}{81}\right\} \\

(2,\tfrac{3}{4}) 
& \left\{3\times\tfrac{2}{3},\,\tfrac{32}{81},\,0.3098\ldots,\,\tfrac{64}{243},\,0.2219\ldots,\,
2\times\tfrac{40}{243}\right\} \\

(2,\tfrac{15}{4}) 
& \left\{\tfrac{2}{3}\right\} \\

(\tfrac{15}{4},0) 
& \left\{1,\,2\times\tfrac{8}{27}\right\} \\

(\tfrac{15}{4},2) 
& \left\{1,\,2\times\tfrac{8}{27}\right\} \\

(6,\tfrac{3}{4}) 
& \left\{\tfrac{32}{81}\right\} \\

(6,\tfrac{15}{4}) 
& \left\{\tfrac{32}{81}\right\}
\end{array}
\]
The eigenvalues \(\frac{2}{3}, \frac{1}{2}, 1,\) and \(\frac{3}{8}\)
are inherited from the deformed Hamiltonian on \(V_{(1)}^0\). This spectrum is consistent with the descendant structure inherited from lower
levels. A primary at \(h=1\), \(j=0\) gives rise to five global descendants at
\(h=2\), \(j=0\), while a primary at \(h=\tfrac{3}{2}\), \(j=\frac{1}{2}\) gives rise to
four global descendants at \(h=2\), \(j=0\). As an illustration, consider the inherited eigenvalue \(\tfrac{2}{3}\). At
\(h=1\), \(j=0\), this eigenvalue is associated with six primary states. Since
each such primary has five global descendants at \(h=2\), \(j=0\), we expect
the eigenvalue \(\tfrac{2}{3}\) to appear with total multiplicity
\[
5\times 6=30
\]
at \(h=2\), \(j=0\). This agrees with the sector-by-sector decomposition. The eigenvalue
\(\tfrac{2}{3}\) appears with multiplicity \(3\) in the sector
\((2,\tfrac{3}{4})\), and with multiplicity \(1\) in the sector
\((2,\tfrac{15}{4})\). After multiplying by the dimensions of the corresponding
\(SU(2)_a\times SU(2)_b\) representations, one obtains
\[
3\times 3\times 2 + 1\times 3\times 4 = 30 .
\]

\section{Casimir sectors used in the level spacing analysis}
\label{app:spacing-data}

In this appendix we list the symmetry-resolved Casimir sectors used in
the main text. In each case, we first restrict to primary
states, remove descendant contributions, unfold the spectrum within each
symmetry-resolved block, and then pool the resulting nearest-neighbor spacings.

\[
\begin{array}{c|c|c}
\text{sector} & \text{regime} & (c_1,c_2)\ \text{Casimir sectors included} \\
\hline
(h,j)=(4,0)
& \textrm{Planar}
& (2,\frac34),\ (0,\frac34),\ (\frac34,2),\ (\frac34,0)
\\[1mm]
(h,j)=(4,0)
& N=3
& (\frac{15}{4},2),\ (\frac34,0),\ (\frac34,2),\ (0,\frac34),
  \ (\frac{15}{4},0),\ (2,\frac34),\ (6,\frac34)
\\[1mm]
(h,j)=(\frac92,\frac12)
&\textrm{Planar}
& (2,0),\ (0,2),\ (0,0)
\\[1mm]
(h,j)=(\frac92,\frac12)
& N=3
& (0,0),\ (0,2),\ (2,0),\ (2,2)
\end{array}
\]
These are the sectors whose unfolded spacings are pooled in the corresponding
histograms in Fig.~\ref{fig:spacing-transition}. The spectra in these sectors are provided in the ancillary file \texttt{data.tex}.

\bibliographystyle{apsrev4-1}
\bibliography{refs}

\end{document}